\documentstyle[a4wide,epsf,11pt,titlepage]{article}

\newcounter{nref}
\newcommand{\bbib}{%
  \renewcommand{\refname}{\large\bf References}%
  \begin{thebibliography}{99}%
  \setcounter{enumiv}{\thenref}}
\newcommand{\ebib}{%
  \end{thebibliography}%
  \setcounter{nref}{\arabic{enumiv}}}
\newcommand{\head}[3]{%
  \setcounter{nref}{0}%
  \thispagestyle{empty}%
  \section*{\LARGE\bf #1}%
  \stepcounter{section}%
  \addcontentsline{toc}{section}{#1}%
  \large\itshape%
  #2\\\vspace{0.1pt}\\%
  #3%
  \normalsize\upshape%
  \bigskip}

\def\msol{\,{\rm M}_\odot}

\def\beq{\begin{equation}}
\def\eeq{\end{equation}}
\def\gcc{\,\,\,{\rm g}\,{\rm cm}^{-3}}
\def\simgr{\,\hbox{\hbox{$ > $}\kern -0.8em \lower 1.0ex\hbox{$\sim$}}\,}
\def\simle{\,\hbox{\hbox{$ < $}\kern -0.8em \lower 1.0ex\hbox{$\sim$}}\,}

\begin{document}


\head{Cooling of neutron stars: effects of accreted envelopes, 
magnetic field and crustal superfluidity}
     {G. Chabrier$^1$, A. Y. Potekhin$^2$ \& D. G. Yakovlev$^2$}
     {$^1$ Ecole Normale Sup\'erieure de Lyon, CRAL (UMR CNRS 5574), Lyon, France\\
      $^2$ Ioffe Physico-Technical Institute, St.-Petersburg, Russia}

\subsection*{Abstract}

We briefly review recent theoretical studies of the effects of
accreted envelopes, magnetic fields and crustal superfluidity on the cooling of
neutron stars.  These effects are especially important for slowly
cooling  low-mass neutron stars, where direct Urca process  of neutrino
emission is forbidden. The effects are useful for interpretation of
observations of several isolated middle-aged neutron stars.

\subsection{Introduction}

It is conventional to separate a neutron star (NS) into the isothermal
interior and the outer heat-blanketing envelope. 
The envelope can be treated
separately in the plane-parallel quasi-Newton\-ian approximation
(Gudmundsson et al.\ \cite{chab.1}; hereafter GPE). 
The evolution of NSs is controlled by the
relationship between the surface temperature $T_s$ and the temperature
$T_b$ at the envelope bottom, at a density $\rho_b \simle
10^{10-11}\,\gcc$. The 
temperature distribution in the blanketing envelope is determined by
the equation (see, e.g., GPE)

\beq
{d T\over d P}={3\over 16}{K \over g}{T_s^4\over T^3}\,,
\eeq

\noindent where $g$ is the surface gravity, $P$ is the pressure, and
$K=1/(K_r^{-1}+K_e^{-1})$ denotes the total mean opacity, including
the radiative
and electron
contributions ($K_r$ and $K_e$, respectively).

The first accurate model of a non-magnetized, pure iron blanketing envelope was
developed by GPE.
This model has
been improved in recent years in several respects.
The thermodynamic and transport properties of the plasma 
in the envelope have been updated; a possible presence of accreted,
light-element surface layers has been taken into account 
\cite{chab.3,chab.4,chab.5,chab.6}; 
the effects of magnetic fields in the blanketing
envelope have been analyzed \cite{chab.14,chab.6}. 
In addition, the effect of superfluidity of
neutrons in inner NS crusts (at densities
from neutron drip, $\sim 4 \times 10^{11}\,\gcc$,
to $\sim 1.5 \times 10^{14}$ g cm$^{-3}$)
on the cooling
has been considered \cite{chab.6,chab.17}.
Other references can be found in \cite{chab.21}.

Here we summarize how the physics
of NS crusts affects the cooling. 
The effect is particularly important for low-mass stars, where  the
powerful direct  Urca process of neutrino emission is forbidden by
momentum conservation law.

Below we present theoretical cooling curves
(effective surface temperatures as detected
by a distant observer, $T_s^\infty$ versus stellar age $t$)
calculated with the physics input
described in Ref.\ \cite{chab.21}.
We use the same version of a moderately stiff
equation of state 
in NS cores as in \cite{chab.21}.
In this model
the direct Urca process
is open in the inner cores of NSs
with gravitational mass
$M > 1.358\msol$.
We compare theoretical results with
observations of thermal radiation from isolated
middle-aged
NSs. The observational
basis is the same as in \cite{chab.21}. 

\subsection{Effect of accreted envelopes}

\begin{figure}[t]
   \centerline{\epsfxsize=0.6\textwidth\epsffile{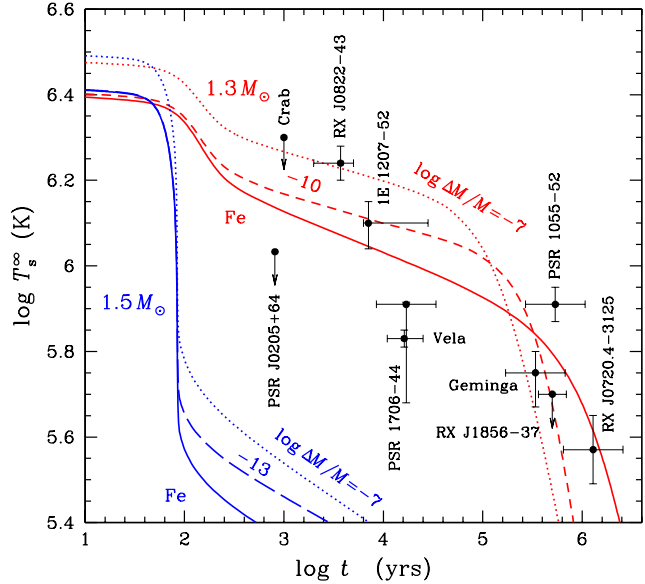}}
  \caption{Cooling of nonsuperfluid 1.3 and 1.5 $\msol$ 
  NSs with different masses $\Delta M$ of 
  accreted layers, as compared with observation. 
  Solid curves refer to nonaccreted (Fe) envelopes.}
  \label{chabfig1.eps}
\end{figure}

Figure \ref{chabfig1.eps} illustrates the effect of accreted layers 
on cooling of 
nonsuperfluid NSs of two masses.  
A massive star with
$M=1.5\,\msol$ gives an example of rapid cooling dominated by direct 
Urca
 process, whereas
a low-mass star, $M=1.3\,\msol$, gives an example of slow 
neutrino cooling 
via modified Urca
process. Drops
of cooling curves at $t \sim 100$ yr
manifest the end of thermal relaxation between stellar
core and crust. Drops
at $t \sim 10^5$--$10^6$ yr indicate the transition
from the neutrino cooling stage to the photon cooling stage. 
Accreted envelopes have
opposite effects
at these two stages. At
the neutrino
cooling stage, the more heat-transparent 
blanketing envelope, due
to accretion of light elements, 
leads to a larger surface temperature than in the case
of iron envelope.
In contrast, during the photon cooling stage, 
the lower thermal insulation of the accreted envelope
leads to a faster cooling. Even a very small fraction of accreted matter
 ($\Delta M/M \sim 10^{-16}$)
changes appreciably the cooling
\cite{chab.3,chab.4,chab.6}. 

\subsection{Effect of magnetic field}

\begin{figure}[t]
   \centerline{\epsfxsize=0.6\textwidth\epsffile{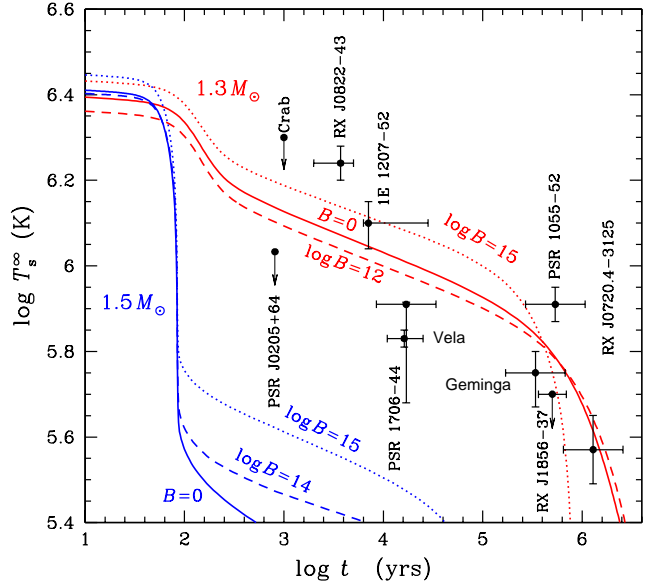}}
  \caption{Same as in Figure 1 for nonaccreted 
  envelopes with several dipole magnetic fields.}
  \label{chabfig2.eps}
\end{figure}

The thermal structure of 
NSs is strongly affected by magnetic fields. 
Landau quantization of electron orbits
in a magnetic field affects
the thermodynamic properties, the heat conduction, 
and the radiative opacities in the surface layers
\cite{chab.9,chab.8,chab.10,chab.19}.
In particular, Landau quantization enhances heat
transport along the field lines and thus increases
$T_s$ (for a given $T_b$) near magnetic poles.
By contrast, classical effect of electron Larmor
rotation reduces electron thermal conductivity
across the field lines, decreasing $T_s$ near magnetic
equator.    
The case of arbitrary
inclination of the magnetic field to the NS
surface
has been studied
in
\cite{chab.11,chab.12,chab.13,chab.15}. 
More recently Potekhin et al.\ \cite{chab.14,chab.6} used improved electron
conductivities of magnetized envelopes \cite{chab.16}, as well as
improved radiative opacities and equation of state of partially ionized,
strongly magnetized hydrogen in 
NS atmospheres \cite{chab.10,chab.19}, to
calculate the cooling of NSs 
with strong dipole magnetic fields.
Figure \ref{chabfig2.eps} displays the effect of dipole magnetic
fields on the cooling of nonsuperfluid high-mass and
low-mass NSs. The curves are marked
by magnetic field strength at the magnetic pole.
At $B \simgr 10^{12}$ G the temperature distribution
over the stellar surface becomes strongly anisotropic,
with the magnetic poles much hotter than the equator.
Figure \ref{chabfig2.eps} shows the effective temperature
$T_s^\infty$ averaged over the surface (see, e.g., \cite{chab.14,chab.6}).
As in the case of 
accreted envelopes, magnetic fields have opposite effects
at the neutrino and photon cooling stages. Fields with $B\simle
10^{13}$ G make the blanketing envelope overall less
heat-transparent, decreasing $T_s$ at the neutrino cooling stage 
and increasing $T_s$ at the photon cooling stage
(especially for low-mass stars). 
Stronger fields have the opposite effect.

\subsection{Effect of crustal superfluidity}

Observational data can be explained assuming
a weak triplet-state pairing of neutrons and a strong
singlet-state pairing of
protons in stellar cores (see, e.g., Ref.\ \cite{chab.21}
and references therein). Let us focus on cooling
of low-mass NSs in the frame of this scenario.
The strong proton superfluidity switches off the modified Urca process
in the cores.
The cooling of these stars becomes exceptionally slow.
It turns out to be
rather insensitive to the equation of state
and models for strong proton
superfluidity in stellar cores. 
However, it becomes sensitive to the models for
singlet-state neutron pairing in stellar crusts (whereas
cooling of more massive stars does not depend on
the crustal superfluidity).

\begin{figure}[t]
   \centerline{\epsfxsize=0.6\textwidth\epsffile{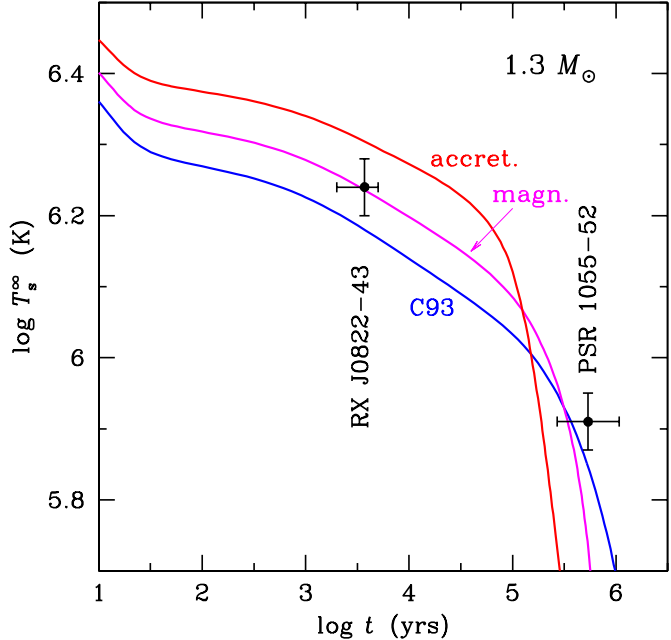}}
  \caption{Cooling of 
  NS with strongly superfluid protons in the core
  and superfluid neutrons in the crust (curve ``C93'').
  Curves ``accret'' and ``magn'' include, in addition,
  the effects of accreted envelope ($\Delta M/M=10^{-7}$)
  and surface magnetic field ($B=10^{15}$ G), respectively. 
}
  \label{chabfig3.eps}
\end{figure}

This effect is illustrated in Fig.\ \ref{chabfig3.eps}
which shows (curve ``C93'') cooling of a star ($M= 1.3\,\msol$)
with a strong superfluidity of protons (model 1p of Ref.\ \cite{chab.21}),
no triplet-state superfluidity of neutrons in the core,
and a model
of Chen et al.\ \cite{chab.18}
for singlet-state crustal neutron superfluidity.
The crustal superfluidity
initiates neutrino emission due to Cooper pairing of
neutrons. Because the neutrino emission from 
the stellar core is weak, the crustal neutrino
emission noticeably accelerates the cooling \cite{chab.6}.
The acceleration is not too strong and enables to explain 
the observations of PSR B1055--52, although violates
interpretation of
the observations of RX J0822--4300. In order to explain
the latter source, one can assume that it
has either a  strong magnetic field 
or an accreted envelope.    

\subsection{Conclusion}

Our calculations show that detailed models of 
NS cooling, 
taking into account the effects of accretion, 
dipole magnetic fields and crustal superfluidity 
can help to explain
observations.
These effects are most important for
slowly
cooling low-mass NSs -- isolated middle-aged NSs hottest for their ages.
Although several models of crustal superfluidity are currently
consistent with observational error bars of effective
temperatures $T_s^\infty$ of these objects, 
better measurements of 
$T_s^\infty$ 
in the future
should lead to a better determination of the surface
magnetic fields and the masses of accreted envelopes of these
objects
and should allow one to discriminate between 
models of crustal superfluidity. 

\subsection*{Acknowledgements}

The work of A.P.\ and D.Y.\ is supported in part by
RFBR (grants 02-02-17668 and 03-07-90200)
and by the Russian Leading Scientific Schools Foundation 
(grant 1115.2003.2).


\end{document}